\documentclass{article}
\usepackage{arxiv}
\usepackage[utf8]{inputenc}
\usepackage[T1]{fontenc}
\usepackage[square,numbers,sort&compress]{natbib}
\usepackage{hyperref}
\usepackage{url}
\usepackage{booktabs}
\usepackage{amsfonts}
\usepackage{microtype}
\usepackage{graphicx}
\usepackage{threeparttable}
\usepackage{multirow}
\usepackage{fontawesome5}
\usepackage{multicol}
\usepackage{xcolor}

\usepackage{subcaption}
\usepackage[font=small,labelfont=bf]{caption}
\usepackage{amsmath,amsfonts,bm}
\usepackage{amssymb,amsthm}
\usepackage{enumitem}
\usepackage{adjustbox}
\usepackage{float}
\usepackage{array}
\usepackage{makecell}
\usepackage{footmisc}
\usepackage{wrapfig}
\usepackage{tcolorbox}
\usepackage{tikz}
\tcbuselibrary{breakable, skins}
\usetikzlibrary{arrows.meta,positioning,fit,calc}
\usepackage{cleveref}
\usepackage{titlesec}
\usepackage{makecell}
\usepackage[table]{xcolor}
\setlist[itemize,1]{leftmargin=\dimexpr 18pt}
\setlist[enumerate,1]{leftmargin=\dimexpr 18pt}
\renewcommand{\arraystretch}{1.08}
\titlespacing*{\section}{0pt}{1.2\baselineskip}{0.7\baselineskip}
\titlespacing*{\subsection}{0pt}{0.9\baselineskip}{0.45\baselineskip}
\titlespacing*{\subsubsection}{0pt}{0.7\baselineskip}{0.35\baselineskip}

\title{
\raisebox{-0.18\height}{\includegraphics[width=0.045\textwidth]{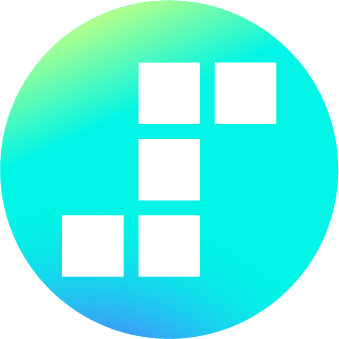}} 
StepAudio 3 Realtime Technical Report
}

\author{\vspace{1em} StepFun-Audio Team}

\renewcommand{\headeright}{StepFun-Audio Team}
\renewcommand{\undertitle}{}
\renewcommand{\shorttitle}{}

\begin{document}
\large

\maketitle

\vspace{-1.2em}

{\centering
\small
\href{https://stepaudiollm.github.io/step-audio-3-realtime/}
{\faHeadphones\ \textbf{Project Page}}
\par}

\vspace{0.6em}

\begin{abstract}
Realtime spoken interaction demands deep reasoning, prompt responses, and fluid turn-taking. We present StepAudio 3 Realtime, an audio-language foundation model organized around a continuous \textit{listen-converse-think-act} loop. Deep Perception captures rich acoustic cues to interpret user intent, while Seamless Duplex models synchronized audio streams to handle pauses, backchannels, and interruptions naturally. Crucially, we resolve the tension between deep deliberation and latency via Think-While-Speaking, executing private reasoning in parallel with spoken delivery. In reasoning mode, StepAudio 3 reaches a 73.0 macro average on StepAudioChat. With Think-While-Speaking, it achieves dialogue and reasoning performance comparable to dedicated reasoning models while speaking in real time. Furthermore, an integrated Voice Agent handles asynchronous tool execution without disrupting the dialogue flow. StepAudio 3 Realtime achieves top-tier performance across key dimensions: an exceptional 90.6 on the MMSU benchmark, 98.9 Overall on the Artificial Analysis Full-Duplex Bench, and a 56.0\% macro task-success rate on $\tau$-Voice.
\end{abstract}

\begin{figure}[H]
    \centering
    \includegraphics[width=0.9\textwidth]{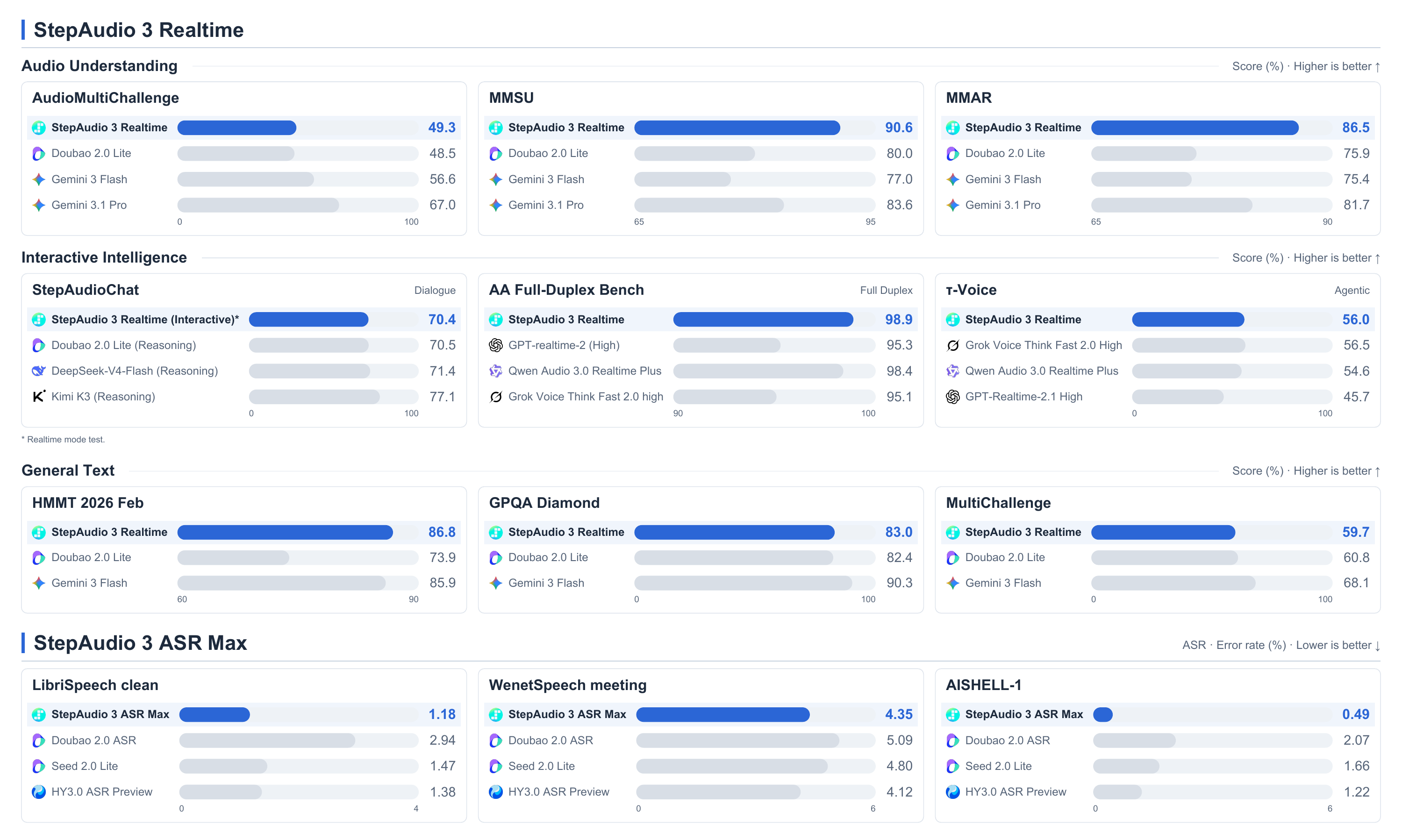}
    \caption{Benchmark results for StepAudio 3 ASR Max and StepAudio 3 Realtime (blue), compared with baselines (gray). Lower is better for ASR error rates, and higher is better elsewhere. Dialogue averages eight dimensions, full duplex reports Overall, and $\tau$-Voice averages three domains. See Tables~\ref{tab:perception:asr_results} and~\ref{tab:overall-bmk}.}
    \label{fig:main-results}
\end{figure}
\clearpage

\section{Introduction}
\label{sec:introduction}

Natural spoken interaction requires a system to follow the user while managing its own response. A pause may occur before a request is complete, and an utterance during model speech may be an acknowledgment or a substantive interruption. Complex requests introduce another challenge: the model must reason carefully while keeping the conversation responsive. Tool use extends this challenge because an external task may outlast the spoken exchange that initiated it.

Advances in speech recognition have combined acoustic representations with the linguistic knowledge of large language models~\citep{radford2023robust,bai2024seed,lin2026paraasr}. Audio-language models now support broader acoustic understanding and direct speech generation~\citep{borsos2023audiolm,tang2024salmonn,kim2024paralinguistics,wu2025step,xu2025qwen3}. Streaming and full-duplex systems further allow listening and speaking to overlap~\citep{wang2024freeze,defossez2024moshi,wu2026chronological,zhang2026duplexsla}. Together, these capabilities allow responses to account for linguistic content, vocal delivery, and conversational timing.

StepAudio 3 Realtime builds on the Step-Audio series' shared audio-language foundation~\citep{huang2025stepaudio,tian2025step,zhang2026step,lin2026stepaudio}. Its focus is the coordination of perception, reasoning, and action as a conversation unfolds. We organize these functions as a listen, converse, think, and act loop. Deep Perception captures linguistic and nonverbal acoustic evidence. Seamless Duplex uses user and model speech to manage the conversational floor. Think-While-Speaking~\citep{wu2025mind} coordinates reasoning with spoken delivery, supported by Adaptive Thinking and multi-token prediction. A streaming Voice Agent carries conversational intent into tool execution and incorporates the results into subsequent dialogue. These functions operate concurrently as needed, with new user input shaping the ongoing interaction.

Figure~\ref{fig:main-results} summarizes ASR results for StepAudio 3 ASR Max and the capability evaluations of StepAudio 3 Realtime. The realtime model leads the reported baselines on four of eight audio-understanding benchmarks and achieves the highest reported Overall score on the Artificial Analysis Full-Duplex Bench. The results also identify remaining gaps in multi-turn constraint following and retail tool-use tasks. Section~\ref{sec:evaluation} presents the protocols and domain-level analysis.

\section{Realtime Conversational Loop}
\label{sec:architecture}

\begin{figure}[!htbp]
\centering
\includegraphics[width=\linewidth]{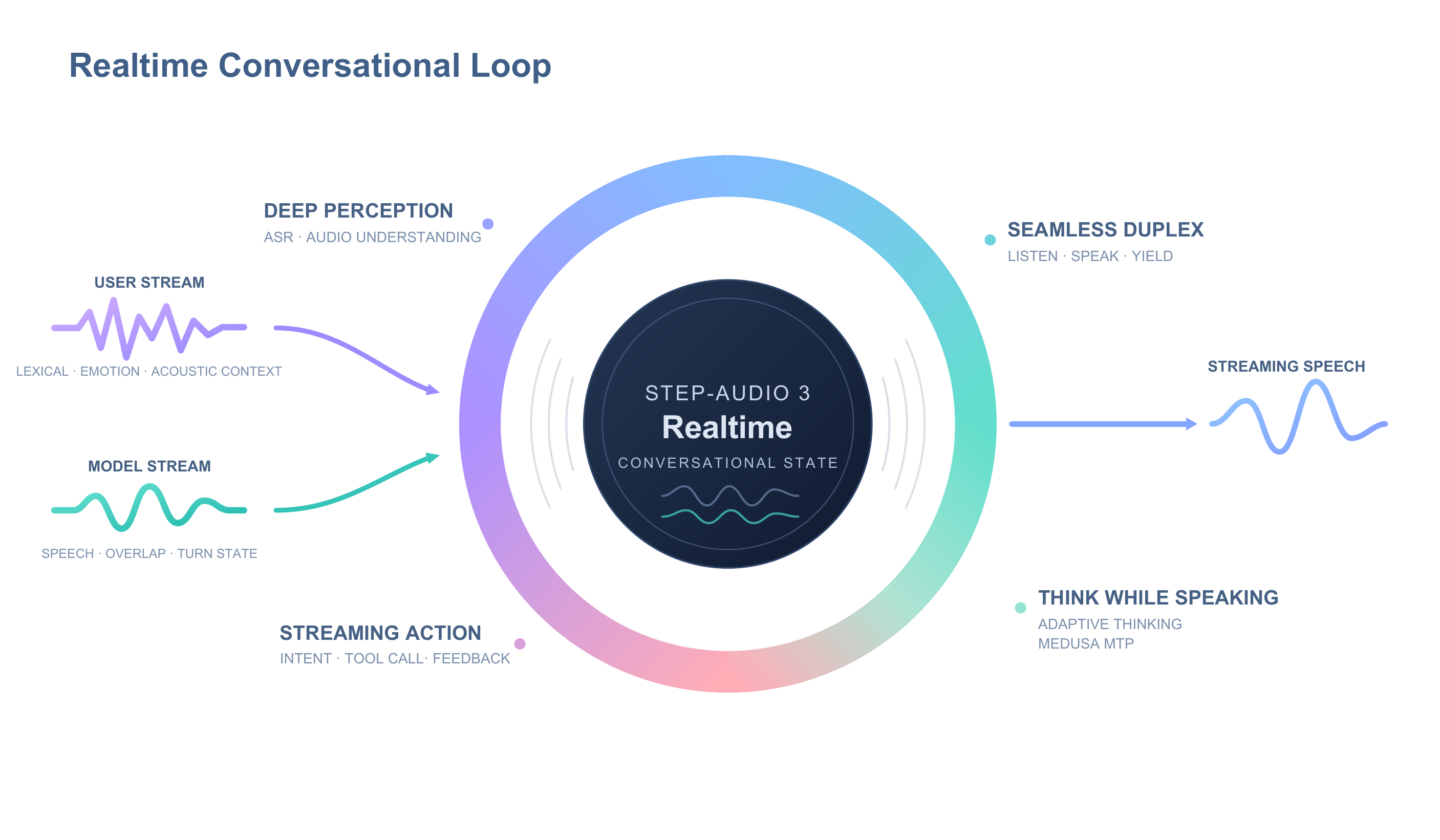}
\caption{Conversational loop of StepAudio 3 Realtime. User and model speech inform perception and floor management. Reasoning supports spoken responses and tool use, while returned tool results update the context for subsequent interaction.}
\label{fig:realtime_conversational_architecture}
\end{figure}

StepAudio 3 Realtime coordinates listening, speaking, reasoning, and action through an evolving conversational context. User speech, model speech, and tool results can arrive while other parts of the interaction remain in progress. Figure~\ref{fig:realtime_conversational_architecture} gives an overview of these coupled functions.

\subsection{Shared Conversational Context}

The conversational context includes acoustic and linguistic evidence, dialogue history, the current speaking turn, reasoning progress, and tool-execution status. Perception retains cues about what the user says and how it is said. Model-side speech provides additional context for interpreting user utterances that overlap with a response.

This context informs whether to continue listening or speaking, whether to reason further, and whether a request is ready for external action. Newly observed speech and returned tool results can change these decisions as the conversation proceeds.

\subsection{Coordinating Speech, Reasoning, and Action}

Conversational timing and reasoning progress need not advance at the same pace. Seamless Duplex handles pauses, user backchannels, and substantive interruptions. Think-While-Speaking allows spoken delivery to begin before the full reasoning trace is complete. Adaptive Thinking selects when explicit reasoning is useful, and MTP accelerates private reasoning. Sections~\ref{sec:duplex} and~\ref{sec:reasoning:mps} describe these capabilities.

The Voice Agent extends the interaction to tasks that require tools. It resolves the request and required arguments before execution, then incorporates returned evidence into the conversation. The user can continue speaking while a task is in progress. This coordination lets external work proceed alongside dialogue, as described in Section~\ref{sec:execution}.

\section{Model Architecture and Foundation Training}
\label{sec:foundation_training}


\subsection{System Architecture}
\label{sec:foundation:architecture}

StepAudio 3 Realtime uses a mixture-of-experts architecture with approximately 196 billion total parameters and 11 billion active parameters per token. Its language backbone is based on Step 3.7 Flash~\citep{stepfun2026step37flash}. The audio frontend uses the Audio Transformer (AuT) encoder from Qwen3-Omni~\citep{xu2025qwen3}. An adapter maps the encoder outputs into the representation space of the language model.

\begin{figure}[!htbp]
\centering
\includegraphics[width=\linewidth]{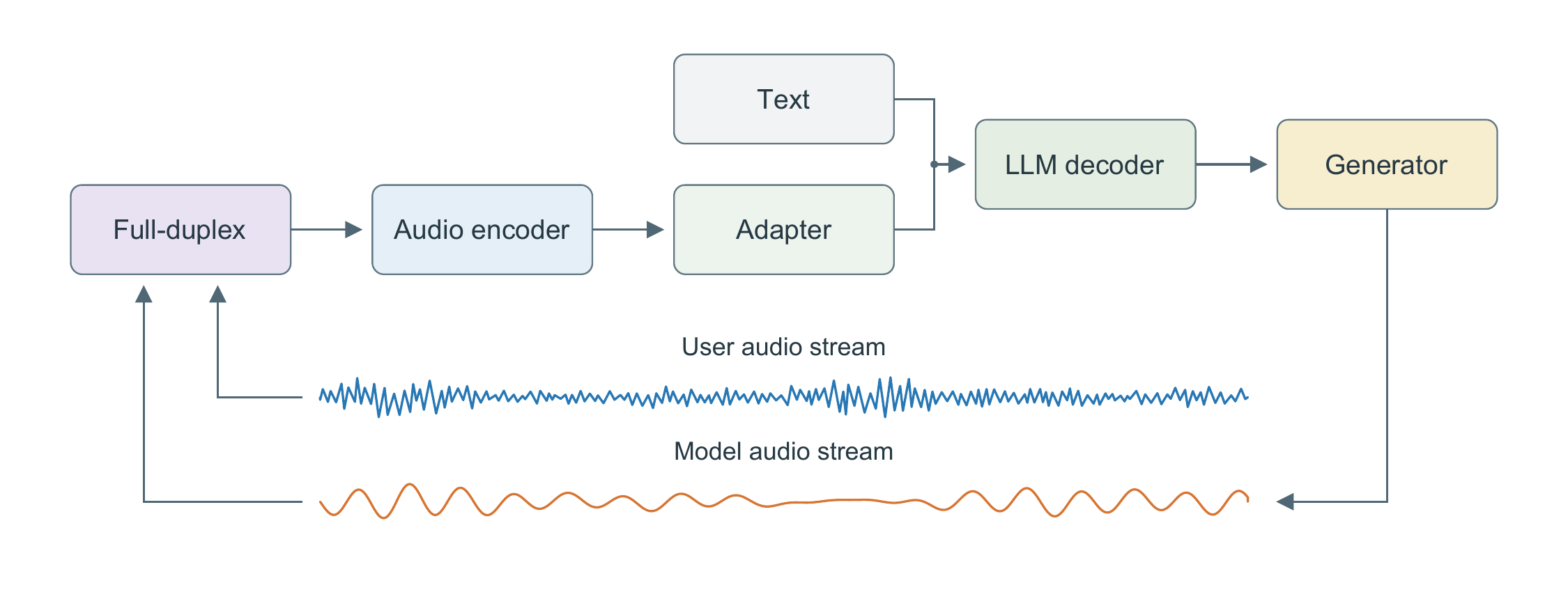}
\caption{System architecture of StepAudio 3 Realtime. The LLM decoder receives audio representations through the audio encoder and adapter, together with a separate text input. User and model audio form the two full-duplex streams, with generator output returning to the model audio stream. Waveforms are schematic.}
\label{fig:streaming_architecture}
\end{figure}

Figure~\ref{fig:streaming_architecture} summarizes the system architecture. The full-duplex input path incorporates user and model audio streams. Audio representations pass through the encoder and adapter to the LLM decoder. Text tokens enter the decoder through a separate input path, allowing it to jointly condition on acoustic information and textual context. The generator produces streaming model audio, which returns to the model audio stream for subsequent interaction. Section~\ref{sec:duplex} describes conversational-floor management.

The speech generator produces incremental output with context-appropriate tone and rhythm. Natural delivery includes expressive cues such as pauses and hesitation, connecting the content of a response with its communicative intent.

\subsection{Three-Stage Pretraining}

\textbf{Data curation.} Pretraining data are prepared through an automated large-scale audio curation pipeline~\citep{lin2026stepaudio}. Raw audio is filtered with sound event detection and voice activity detection, then merged and resegmented into samples of suitable duration that preserve semantic completeness. The pipeline assigns audio-level metadata such as quality, synthetic-speech likelihood, and speaker count. It also uses multiple recognition systems for transcription and language identification, cross-checks their outputs, and grades samples by acoustic and semantic quality. These annotations support quality-aware sampling across training stages. For StepAudio 3 Realtime, the pipeline is extended to broaden language coverage and support the sustained perception and interaction demands of realtime dialogue.

\textbf{Training stages.} Pretraining is organized into modality alignment, multimodal mixed training, and cooldown stages. The modality-alignment stage establishes the interface between acoustic representations and the language model. Multimodal mixed training then develops joint audio-text modeling at scale. The cooldown stage places greater weight on high-quality data to refine the resulting foundation. Across the three stages, StepAudio 3 Realtime uses a fixed sequence length of 32K and processes 1.2T training tokens.

\textbf{Pretraining mixture.} The pretraining mixture increases the proportion of pure text to preserve the general capabilities of the base language model and support subsequent reasoning and agent training.

\subsection{Midtraining for Realtime Interaction}

\textbf{Context extension.} Midtraining uses perception, synthetic conversational, and voice-agent data. The context length is extended to 128K to accommodate longer dialogue histories, earlier user requirements, and intermediate tool results.

\textbf{Midtraining mixture.} This stage substantially increases the share of audio-understanding and agent-interaction data. The former broadens coverage of speech, music, environmental sound, and audio-grounded reasoning, while the latter trains the model to carry user intent through planning, tool use, and spoken follow-up. Sections~\ref{sec:perception:audio_understanding} and~\ref{sec:execution} describe the corresponding data construction and training procedures.

\section{Deep Perception: Speech Recognition and Audio Understanding}
\label{sec:perception}

Perception combines lexical understanding with cues about the speaker, vocal delivery, acoustic events, and temporal structure. StepAudio 3 ASR Max is specialized for transcription, while StepAudio 3 Realtime is trained for broader audio understanding and spoken interaction. We describe the ASR specialization first, followed by audio-understanding data construction and post-training. Section~\ref{sec:evaluation} compares capabilities across benchmarks.

\subsection{StepAudio 3 ASR Max}
\label{sec:perception:training}

\subsubsection{Training and Data Construction}

StepAudio 3 ASR Max and StepAudio 3 Realtime share the same pretraining and
midtraining stages. They diverge only during supervised fine-tuning, where the
ASR branch is specialized for transcription and the realtime branch is tuned
for spoken interaction.

\textbf{Supervised fine-tuning.} The ASR-specialized model is fine-tuned with examples packed into sequences of up to 32K tokens. We apply time-frequency masking following the augmentation principle of SpecAugment~\citep{park2019specaugment}, while keeping the audio encoder frozen and updating the audio-language adapter and language decoder to produce normalized transcripts. For context-aware recognition, an example may additionally provide dialogue history, a preceding model response, a scenario description, or task-specific terminology as optional evidence. The target transcript remains grounded in the input waveform, allowing the model to use relevant context without simply copying unrelated terms.

\label{sec:perception:asr_data}
\label{sec:perception:data}

\textbf{Short- and long-form ASR data.} The ASR mixture combines short labeled utterances with long pseudo-labeled recordings. Multiple recognition systems transcribe segmented audio, and their hypotheses are aligned and fused with Recognizer Output Voting Error Reduction (ROVER)~\citep{fiscus1997post}. Agreement-based filtering selects reliable segments for recomposition into longer sessions. LLM then restores punctuation and improves consistency across each session.

\textbf{Long-tail terminology augmentation.} Rare names and technical terms are
often confused with common words that sound similar. We therefore build targeted
synthetic training examples for these cases. An LLM expands a knowledge taxonomy
to identify categories rich in homophones, uncommon characters, abbreviations,
and product identifiers. We enumerate candidate terms, remove duplicates, and
place the terms in natural carrier sentences. These sentences are converted to
speech and retained only when their pronunciation is consistent with the target
text.\footnote{For related model- and data-centric methods, including synthetic speech augmentation for code-switching ASR, see~\citep{liu2026codeswitching}.}
For acoustically confusable terms, selected examples may also include dialogue
history or entity hints. This teaches the model to use relevant context while
avoiding unrelated lexical substitutions.

\subsubsection{Evaluation}
\label{sec:perception:evaluation}
\label{sec:perception:evaluation:asr}

\textbf{Benchmarks.} We evaluate StepAudio 3 ASR Max on five standard public test sets: LibriSpeech
test-clean and test-other~\citep{panayotov2015librispeech}, AISHELL-1
~\citep{bu2017aishell}, and WenetSpeech test-net and test-meeting
~\citep{zhang2022wenetspeech}. English results use word error rate (WER), while
Mandarin results use character error rate (CER). To assess the linguistic
knowledge targeted by our long-tail terminology augmentation, we additionally
use the publicly released ContextASR-Bench~\citep{wang2025asrbench}, which
provides long-form, multi-domain, entity-rich speech in English and Mandarin.
We use its Contextless setting without domain labels, entity lists, or external
hotword injection; English subsets are evaluated with WER and Mandarin subsets
with CER.

\begin{table*}[t]
\centering
\small
\setlength{\tabcolsep}{5pt}
\renewcommand{\arraystretch}{1.06}
\caption{ASR evaluation on standard public benchmarks and ContextASR-Bench
(lower is better). English subsets report WER and Mandarin subsets report CER.
Bold marks the best result in each row, and underlining marks the second-best
result. ContextASR-Bench results use the Contextless setting without external
context injection.}
\label{tab:perception:asr_results}
\begin{adjustbox}{max width=\textwidth}
\begin{tabular}{lc|ccc}
\toprule
\textbf{Test set} &
\textbf{StepAudio 3 ASR Max} &
\textbf{Doubao 2.0 ASR} &
\textbf{Seed 2.0 Lite} &
\textbf{HY3.0 ASR Preview} \\
\midrule
LibriSpeech test-clean
& \textbf{1.18} & 2.94 & 1.47 & \underline{1.38} \\
LibriSpeech test-other
& \textbf{2.28} & 5.98 & \underline{2.67} & 2.80 \\
AISHELL-1
& \textbf{0.49} & 2.07 & 1.66 & \underline{1.22} \\
WenetSpeech test-net
& \underline{3.99} & 4.03 & 4.71 & \textbf{3.71} \\
WenetSpeech test-meeting
& \underline{4.35} & 5.09 & 4.80 & \textbf{4.12} \\
\midrule
\multicolumn{5}{l}{\textit{ContextASR-Bench}} \\
\hspace{1em}- ContextASR-Speech-EN
& \textbf{7.91} & 12.04 & 9.48 & \underline{8.53} \\
\hspace{1em}- ContextASR-Dialogue-EN
& \textbf{3.43} & 9.09 & \underline{3.65} & 4.66 \\
\hspace{1em}- ContextASR-Speech-ZH
& \textbf{1.43} & 2.80 & 2.15 & \underline{1.74} \\
\hspace{1em}- ContextASR-Dialogue-ZH
& \textbf{1.02} & 10.47 & 4.15 & \underline{1.63} \\
\bottomrule
\end{tabular}
\end{adjustbox}
\end{table*}

\textbf{Results.} Table~\ref{tab:perception:asr_results} compares ASR performance across benchmark subsets. StepAudio 3 ASR Max leads on all three standard benchmark families: it is best on both LibriSpeech subsets and AISHELL-1, and it remains ahead of Doubao 2.0 ASR and Seed 2.0 Lite on both WenetSpeech subsets while trailing HY3.0 ASR Preview only by a small margin. On ContextASR-Bench, StepAudio 3 ASR Max is the best-performing model on all four subsets, covering both English and Mandarin and both Speech and Dialogue settings. Its macro-average error rate is 5.67\% on the English subsets and 1.23\% on the Mandarin subsets, compared with 6.60\% and 1.69\% for HY3.0 ASR Preview. This consistent lead indicates strong overall transcription accuracy on the benchmark's long-form, multi-domain, entity-rich speech without external context injection. Note that these results characterize the ASR-specialized model, not the transcription behavior of the realtime model.

\subsection{Audio Understanding}
\label{sec:perception:audio_understanding}

\begin{figure}[!htbp]
\centering
\includegraphics[width=0.92\linewidth]{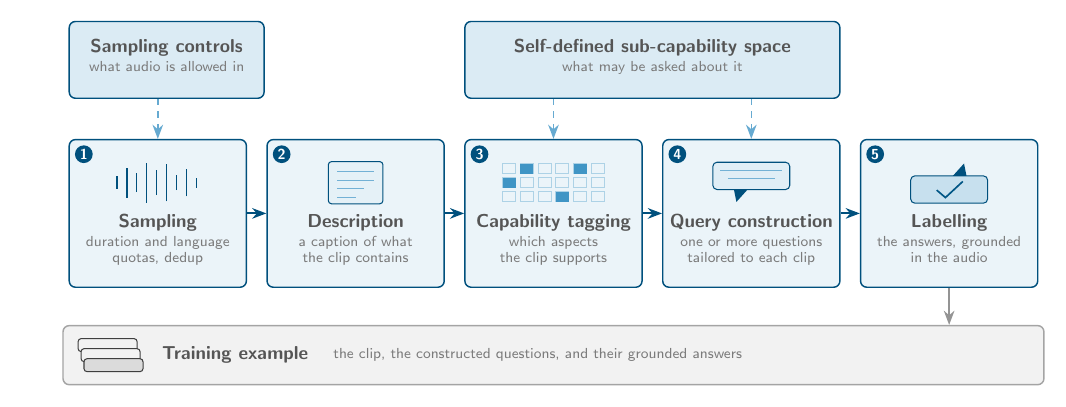}
\caption{Construction pipeline for audio-understanding training examples. Sampling controls select suitable recordings. Description and capability tagging identify what each clip contains and which aspects it supports. Query construction, multi-model labeling, and agreement filtering then produce candidate training examples.}
\label{fig:audio_understanding_data_construction}
\end{figure}

\textbf{Data construction.} A hierarchical taxonomy covers lexical content, paralinguistics, acoustic events, speaker and temporal structure, music, and audio-grounded reasoning. Sampling controls balance duration and language coverage while removing duplicates and unsuitable recordings. Each recording is first described and mapped to the capabilities supported by its content. These annotations guide the construction of one or more clip-specific questions. Multiple models then independently label each audio--question pair, and their outputs are consolidated through agreement and quality checks. Figure~\ref{fig:audio_understanding_data_construction} summarizes the construction process.

\textbf{Data selection and quality control.} Deterministic checks first remove empty, truncated, malformed, or severely repetitive outputs. Text-only LLM judges then assess query and response quality and assign a case-value score. Case value jointly considers the query, the response, the amount of useful information available in the audio as represented by its annotations, and the training value of the question. These judges do not directly evaluate audio grounding. Instead, grounding reliability is estimated from the consistency of responses independently produced by multiple models for the same audio--question pair. Only candidates with high quality, high case value, and strong cross-model consistency are retained as SFT candidates; broadly useful examples may enter midtraining, while disagreements and correctable cases are routed to relabeling or further review.\footnote{Related work examines modality-grounded evaluation and staged post-training in omni-modal models~\citep{liu2026boosting}.}


\textbf{Evaluation.} We evaluate the post-trained conversational model on eight
audio-understanding benchmarks covering audio-grounded reasoning, fine-grained
perception, nonverbal acoustic cues, and multi-turn understanding. Results are
reported in Table~\ref{tab:perception:audio-understanding-results}; the complete
cross-capability comparison and evaluation protocols are provided in
Section~\ref{sec:evaluation}.

\begin{table*}[t]
\centering
\small
\setlength{\tabcolsep}{5pt}
\renewcommand{\arraystretch}{1.06}
\caption{Audio-understanding results on eight benchmarks and their unweighted
macro average. Scores use a 0--100 scale (higher is better). Bold marks the best
result in each row, and underlining marks the second-best result.}
\label{tab:perception:audio-understanding-results}
\begin{adjustbox}{max width=\textwidth}
\begin{tabular}{lc|ccc}
\toprule
\textbf{Benchmark} &
\textbf{StepAudio 3 Realtime} &
\textbf{Doubao 2.0 Lite} &
\textbf{Gemini 3 Flash} &
\textbf{Gemini 3.1 Pro} \\
\midrule
Big Bench Audio
& 98.1 & 98.8 & \underline{99.4} & \textbf{99.6} \\
AudioMultiChallenge
& 49.3 & 48.5 & \underline{56.6} & \textbf{67.0} \\
MMSU
& \textbf{90.6} & 80.0 & 77.0 & \underline{83.6} \\
MMAU
& \underline{79.0} & 77.5 & 77.6 & \textbf{80.5} \\
WildSpeech
& \underline{77.1} & 73.9 & 74.4 & \textbf{77.7} \\
MMAR
& \textbf{86.5} & 75.9 & 75.4 & \underline{81.7} \\
Step-Caption
& \textbf{78.2} & \underline{76.8} & 67.8 & 74.8 \\
MTalk-Bench
& \textbf{72.8} & 66.8 & 66.7 & \underline{70.6} \\
\midrule
\textbf{Macro Average}
& \underline{79.0} & 74.8 & 74.4 & \textbf{79.4} \\
\bottomrule
\end{tabular}
\end{adjustbox}
\end{table*}

StepAudio 3 Realtime leads the reported baselines on four of the eight
benchmarks, with its largest margins on MMSU (90.6 versus 83.6) and MMAR
(86.5 versus 81.7), gains of 7.0 and 4.8 points. It also leads on Step-Caption
and MTalk-Bench, and is close to Gemini 3.1 Pro on MMAU and WildSpeech. In
contrast, it trails Gemini 3.1 Pro by 17.7 points on AudioMultiChallenge, while
Big Bench Audio is nearly saturated for all systems. The results indicate broad
strength in spoken-language understanding, audio-grounded reasoning, and
nonverbal acoustic perception, with maintaining and revising constraints over
natural multi-turn audio remaining a clear area for improvement.

\textbf{Less is more.} We conduct a separate ablation in which the SFT data are
the only changed factor, comparing roughly two million randomly sampled examples
with about 100K high-quality examples retained after quality control. The
quality-controlled set improves MMSU from 78.78 to 89.70 and MMAR from 74.70 to
84.50. WildSpeech rises from 74.20 to 77.11, while the macro average across the
ambient, paralinguistic, and semantic subsets of MTalk-Bench increases from
71.73 to 71.80. Despite using roughly one twentieth as many examples, the
quality-controlled data yield consistent gains, highlighting the importance of
data quality over raw SFT volume.

\section{Seamless Duplex: Conversational Floor Management}
\label{sec:duplex}


A central challenge in full-duplex dialogue is determining when to take, retain, or yield the conversational floor.
This requires distinguishing pauses within an unfinished utterance from turn completion, and brief acknowledgments from attempts to interrupt.
Resolving these ambiguities relies on acoustic evidence interpreted in the context of the unfolding dialogue~\citep{defossez2024moshi,wu2026chronological,fang2026bayling}.

\begin{figure}[H]
    \centering
    \includegraphics[width=\textwidth]{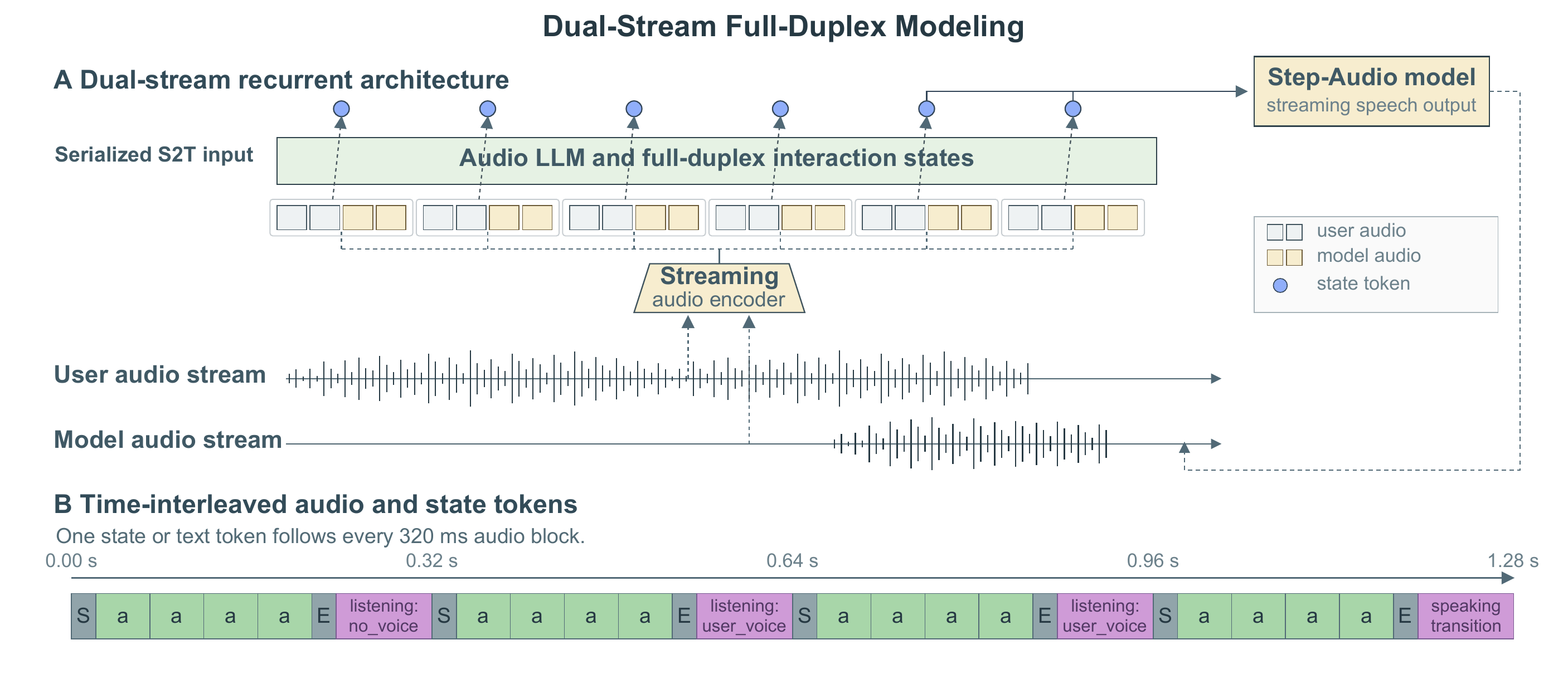}
    \caption{Full-duplex interaction and temporal organization.
    (A) User speech, model-side speech, and dialogue history jointly inform conversational-floor management.
    (B) Audio and interaction states are represented along a shared timeline.}
    \label{fig:duplex-system}
\end{figure}

To support these decisions, StepAudio 3 Realtime integrates incoming user speech, ongoing model speech, and dialogue history for context-aware conversational-floor management.
Figure~\ref{fig:duplex-system} illustrates the dual-stream architecture and the temporal interleaving of audio blocks with interaction-state tokens.


\subsection{Streaming Interaction States}
\label{sec:duplex:state}

The system tracks its interaction state incrementally through the time-interleaved representation illustrated in Figure~\ref{fig:duplex-system}.
Audio is organized into 320\,ms blocks, each followed by a state or text token.
Acoustic evidence and semantic context guide decisions to continue listening, initiate a response, continue speaking, or yield the conversational floor.
The model also uses its own ongoing speech to interpret overlapping user utterances in the context of what the user is currently hearing.






\subsection{Context-Aware Interaction Control}
\label{sec:duplex:dualstream}

The conversational role of a user utterance depends on its relation to the ongoing dialogue~\citep{zhang2026duplexsla}.
For example, ``right'' may function as a backchannel acknowledging the model's explanation or as a preface to a correction.
The system uses both audio streams and dialogue history to interpret these context-dependent roles and guide floor-management decisions.

\paragraph{Pauses and turn endings.}
The system combines acoustic timing with semantic completeness to distinguish within-turn pauses from turn endings.
This distinction guides whether to continue listening or initiate a response.

\paragraph{Backchannels and interruptions.}
During model speech, user backchannels are interpreted in relation to the ongoing response.
Brief acknowledgments can signal continued engagement without requesting a floor transfer, whereas a substantive request or correction may signal an intent to interrupt.
The system uses this distinction to guide whether to continue speaking or yield to the user.

\paragraph{Background speech rejection.}
Dialogue history provides contextual evidence for assessing whether incoming speech is directed at the assistant.
This assessment informs whether the speech should be incorporated into the active exchange or treated as unrelated background conversation.

\subsection{Training}
\label{sec:duplex:training}

\paragraph{Midtraining.}
Midtraining adapts the model to the time-interleaved representation used for full-duplex interaction.
This stage combines supervision for streaming ASR, voice activity detection (VAD), and streaming prediction of utterance completeness.
The training mixture includes over 10,000 hours of synthetic full-duplex interaction data.
Text data are also incorporated to help retain general language and reasoning capabilities.

\paragraph{Post-training.}
Post-training further refines conversational behavior using high-quality interaction data covering turn taking, user backchannel handling, interruption handling, and background speech rejection.

\begin{table*}[t]
\centering
\small
\setlength{\tabcolsep}{5pt}
\renewcommand{\arraystretch}{1.06}
\caption{Full-duplex interaction results on the Artificial Analysis subset of
Full Duplex Bench v1 and v1.5. Scores use a 0--100 scale (higher is better).
Bold marks the best result in each row, and underlining marks the
second-best result.}
\label{tab:full-duplex}
\begin{adjustbox}{max width=\textwidth}
\begin{tabular}{lc|ccc}
\toprule
\textbf{Capability} &
\textbf{StepAudio 3 Realtime} &
\textbf{GPT-realtime-2 (High)} &
\textbf{Qwen Audio 3.0 Realtime Plus} &
\textbf{Grok Voice Think Fast 2.0 High} \\
\midrule
Pause Handling
& \underline{98.9} & \textbf{99.3} & 98.0 & 98.0 \\
Turn Taking
& \textbf{100.0} & \textbf{100.0} & \underline{98.0} & 91.0 \\
User Interruption Handling
& \textbf{99.0} & 95.0 & \underline{98.0} & 97.0 \\
Backchannel Handling
& \underline{98.0} & 86.7 & \textbf{100.0} & 95.0 \\
\midrule
\textbf{Overall}
& \textbf{98.9} & 95.3 & \underline{98.4} & 95.1 \\
\bottomrule
\end{tabular}
\end{adjustbox}
\end{table*}


\subsection{Evaluation}
\label{sec:duplex:evaluation}

StepAudio 3 Realtime ranks first in the Artificial Analysis (AA) full-duplex evaluation, achieving an overall score of 98.9.
This evaluation uses a subset of Full Duplex Bench v1 and v1.5 to assess four aspects of conversational interaction: pause handling, turn taking, user interruption handling, and backchannel handling.

As shown in Table~\ref{tab:full-duplex}, StepAudio 3 Realtime surpasses the strongest baseline, Qwen Audio 3.0 Realtime Plus, which scores 98.4 overall.
Across individual categories, StepAudio 3 Realtime achieves 100.0 on turn taking and 99.0 on user interruption handling, alongside scores of 98.9 on pause handling and 98.0 on backchannel handling.

The category-level results highlight two complementary aspects of full-duplex interaction: respecting within-turn pauses while responding at turn completion, and accommodating user interruptions while continuing through backchannels.
Strong performance across both pairs indicates balanced conversational control over when to listen, speak, and yield.
\section{Conversational Intelligence and Realtime Reasoning}
\label{sec:reasoning}

Seamless Duplex determines when the model should respond. Conversational
intelligence determines how it should engage with the user and how much
reasoning the response requires. A natural voice assistant should follow intent
across turns, clarify underspecified goals, and move the conversation toward a
useful outcome. Routine turns should avoid unnecessary deliberation, while
complex requests should retain the reasoning needed for a reliable answer.

StepAudio 3 Realtime combines dialogue and reasoning training with
Think-While-Speaking, which coordinates spoken responses with ongoing private
reasoning. Adaptive Thinking controls whether a turn uses explicit reasoning,
and MTP acceleration reduces the decoding cost of that reasoning.

\subsection{StepAudioChat Benchmark}
\label{sec:reasoning:benchmark}

StepAudioChat is a closed, text-based benchmark for foundational conversational
intelligence. It evaluates dialogue behavior and the reasoning expressed through
dialogue. Its scope isolates text-level response quality from prosody, turn
timing, interruption handling, and other properties of the speech interface.\footnote{For complementary evaluation of emotional intelligence in multi-turn spoken dialogue, see Multi-Bench~\citep{deng2025multi}.}
The benchmark provides a common capability taxonomy and newly constructed items
to reduce reliance on public test questions.

\subsubsection{Capability Taxonomy}
\label{sec:reasoning:benchmark:taxonomy}

We organize conversational abilities into a hierarchy whose leaves target
observable behaviors with defined evaluation boundaries. Tasks and
representative examples from public benchmarks serve as a coverage check,
without reusing their test questions. Overlapping mappings help refine
capability definitions, while unmapped examples identify coverage gaps.

Within each capability family, we also vary factors such as the source of a
constraint, its form of expression, and its interaction with other conditions.
This exposes combinations including nested constraints and requirements that
must remain consistent across turns. Each item targets a primary capability.
Only capabilities supported by validated items enter the evaluation suite.

\subsubsection{Naturalistic Item Construction}
\label{sec:reasoning:benchmark:construction}

Each item contains a dialogue prompt, independently checkable criteria, a valid
reference response, and a deliberately flawed response. The paired responses
provide positive and negative controls for the judging criteria while allowing
multiple valid phrasings.

De-identified utterances from real interactions inform prompt phrasing and local
context, preserving brevity, fragmentation, colloquial wording, and transcription
noise. Personal entities are replaced with typed placeholders. These utterances
do not supply expected answers or grading criteria. Except for intrinsically
domain-specific capabilities, scenarios are recast across everyday settings to
broaden coverage beyond individual applications.

\subsubsection{Quality Control and Difficulty Calibration}
\label{sec:reasoning:benchmark:quality}

Structural validation checks that prompts are complete and self-contained. The
valid response must satisfy every judging criterion, and the flawed response
must violate at least one. A semantic audit then checks the premise, reference
response, and judging logic. Automated checks provide broad coverage, with
targeted human review of semantic failures and anomalous cases.

Difficulty is calibrated using responses from two reference systems with
different capability levels. Their observed successes and failures guide the
mixture of baseline, discriminative, and difficult examples. The judge is
calibrated separately using the valid and flawed controls and a trusted labeled
sample. These checks help distinguish capability demands from ambiguous wording
or inconsistent grading.

\subsection{Multi-Turn Dialogue Data}
\label{sec:reasoning:dialogue-data}
\label{sec:reasoning:humanness}

\begin{figure}[!htbp]
\centering
\includegraphics[width=0.92\linewidth]{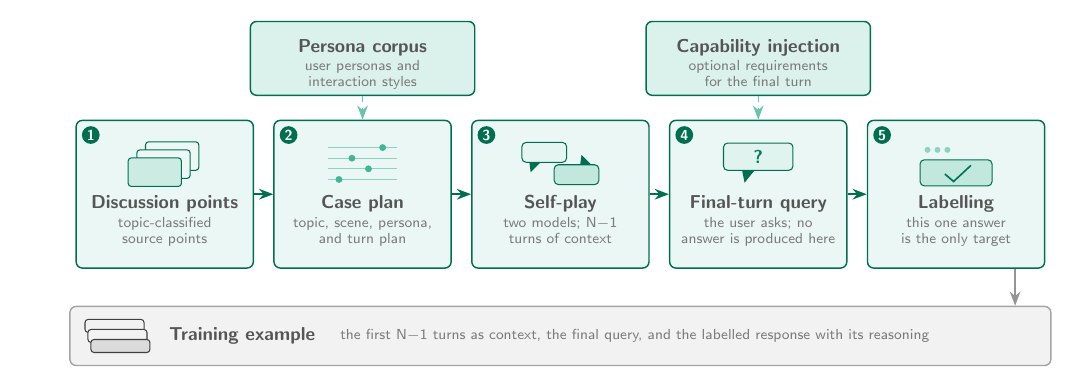}
\caption{Construction of multi-turn dialogue training examples through discussion-point selection, case planning, self-play, final-turn query construction, and labeling.}
\label{fig:multi_turn_dialogue_data}
\end{figure}

We train StepAudio 3 Realtime to follow user intent and constraints across turns,
clarify underspecified requests, and adapt responses to the conversational
context. Joint training on dialogue and reasoning examples supports deliberation
on difficult requests and concise responses to routine turns. Training examples
span diverse topics, personas, and interaction lengths. Open-ended requests can
admit multiple valid responses without requiring a single canonical wording.
Figure~\ref{fig:multi_turn_dialogue_data} summarizes the construction pipeline
for multi-turn dialogue examples that support these objectives.

\paragraph{Factorized construction.}
Generation follows four axes: topics and their concrete discussion points;
participant personas that specify roles, backgrounds, and interaction styles;
turn depth; and target capabilities such as contextual recall, logical reasoning,
and instruction use. A stratified
rotation schedule coordinates these choices within topics, extending coverage
beyond their global marginal distributions.

Turn depth is an explicit construction variable. Longer dialogues progress
through deeper engagement with discussion points, allowing later turns to refine
constraints, resolve ambiguity, revisit evidence, or change direction while
remaining consistent with the history.

\paragraph{Quality assessment and routing.}
Quality control separates three concerns into independent stages. First, a
context--query review evaluates the depth and informativeness of the dialogue
history, whether the opening is self-contained, and whether the final query is
substantive and properly connected to the preceding conversation. Second, when
capabilities are injected into the final turn, a capability-specific review
checks both whether the query instantiates each requested capability and whether
the response actually satisfies its requirement. Third, response review
evaluates answer quality, persona and style consistency, and naturalness as
spoken dialogue.

Deterministic checks handle defects such as empty output, malformed tokens,
severe repetition, and role confusion. Only examples that pass these checks and
receive high scores in all three quality-control stages are retained for
supervised fine-tuning; all other examples are rejected.

\paragraph{Reasoning-model evaluation.}
We evaluate StepAudio 3 in reasoning mode on the eight StepAudioChat
dimensions. This evaluation isolates the model's conversational and reasoning
capability; here we focus only on its dialogue results. The results in this
subsection directly evaluate the reasoning-mode model. By contrast, the overall comparison
in Section~\ref{sec:evaluation} evaluates the realtime system, which additionally
applies Think-While-Speaking, Adaptive Thinking, and MTP acceleration, as
described in Section~\ref{sec:reasoning:mps}. Table~\ref{tab:reasoning:dialogue-results}
compares the reasoning checkpoint with the dialogue baselines.

\begin{table*}[t]
\centering
\small
\setlength{\tabcolsep}{5pt}
\renewcommand{\arraystretch}{1.06}
\caption{Dialogue and reasoning evaluation of StepAudio 3 Realtime in reasoning mode and
baseline models. The reasoning-mode results isolate the model capability before
realtime interaction; the complete realtime system is evaluated separately in
Section~\ref{sec:evaluation}. Scores use a 0--100 scale (higher is better). Bold
marks the best result in each row, and underlining marks the second-best result.}
\label{tab:reasoning:dialogue-results}
\begin{adjustbox}{max width=\textwidth}
\begin{tabular}{lrrrr}
\toprule
Capability & StepAudio 3 Realtime (Reasoning) & Doubao 2.0 Lite & DeepSeek-V4-Flash & Kimi K3 \\
\midrule
Instruction Following & 66.3 & \textbf{72.9} & \underline{71.4} & 68.9 \\
Faithfulness & 72.4 & 67.5 & \underline{75.3} & \textbf{78.4} \\
Reasoning & \underline{73.0} & 72.7 & 64.8 & \textbf{81.9} \\
Memory & \underline{72.0} & 71.3 & 71.5 & \textbf{77.6} \\
Knowledge & \underline{73.1} & 59.9 & 71.6 & \textbf{78.6} \\
Safety \& Reliability & 79.0 & 75.9 & \underline{79.9} & \textbf{84.8} \\
Conversational Pragmatics & \underline{67.2} & 61.5 & 62.9 & \textbf{70.3} \\
Persona \& Role Consistency & \underline{80.9} & \textbf{82.6} & 73.6 & 76.5 \\
\midrule
\textbf{Macro Average} & \underline{73.0} & 70.5 & 71.4 & \textbf{77.1} \\
\bottomrule
\end{tabular}
\end{adjustbox}
\end{table*}

StepAudio 3 achieves a macro average of 73.0, above Doubao 2.0 Lite
at 70.5 and DeepSeek-V4-Flash at 71.4, but below Kimi K3 at 77.1. It ranks
second on reasoning, memory, knowledge, conversational pragmatics, and persona
and role consistency. Kimi K3 leads six of the eight dimensions, including
reasoning at 81.9, memory at 77.6, and safety and reliability at 84.8, while
Doubao 2.0 Lite leads instruction following and persona and role consistency.
These category-level differences show that strong aggregate reasoning does not
imply uniformly stronger instruction following or role consistency.

\subsection{Think-While-Speaking}
\label{sec:reasoning:mps}

Think-While-Speaking coordinates realtime reasoning in StepAudio 3 Realtime.
It builds on the two-process design of Mind-Paced Speaking~\citep{wu2025mind}.
Two concurrent calls to the
same audio model act as a Formulation Brain and an Articulation Brain. The
Formulation Brain generates a private reasoning trace. The Articulation Brain
produces short response segments conditioned on the reasoning available so far
and on the response already spoken.

Playback-aware scheduling releases response segments according to the progress
of the streaming output audio while formulation continues in parallel. Once
formulation finishes, the remaining response can use the complete reasoning
state. A final continuation can supplement or correct an answer that began from
incomplete reasoning. The system uses Speak-First by default, starting the
response without waiting for an initial reasoning prefix. Think-First waits for
a short reasoning prefix before beginning the response.

Adaptive Thinking controls whether a turn uses explicit reasoning. For turns
that do, MTP acceleration speeds up the private thinking stream, following
speculative and multi-token decoding approaches that reduce sequential
target-model steps~\citep{chen2023accelerating,leviathan2023fast,cai2024medusa,gloeckle2024better}.
Together, these mechanisms coordinate reasoning effort, decoding efficiency,
and spoken delivery.

\subsubsection{Adaptive Thinking}
\label{sec:reasoning:adaptive}

\begin{figure*}[t]
    \centering
    \includegraphics[width=\textwidth]{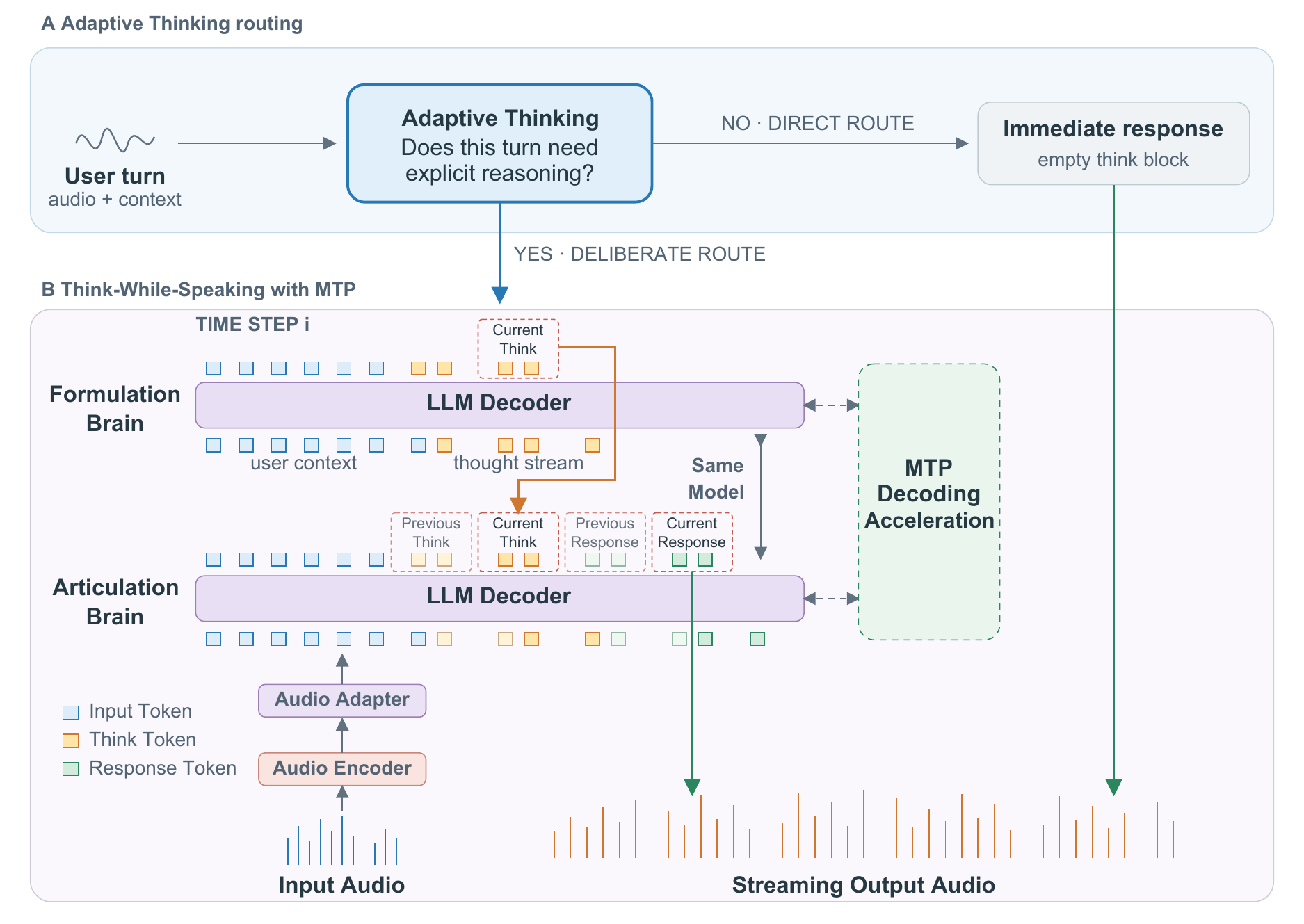}
    \caption{Overview of realtime thinking in StepAudio 3 Realtime. Adaptive Thinking routes each turn to an immediate or deliberate response, while MTP acceleration speeds up the private thinking used by deliberate turns.}
    \label{fig:adaptive-reasoning-overview}
\end{figure*}

\paragraph{Motivation.}
Different conversational turns warrant different deliberation budgets. Explicit
reasoning consumes decoding compute and can delay a routine response. Adaptive
Thinking aims to reduce unnecessary reasoning while retaining it on turns that
benefit from deliberation.

\paragraph{Reasoning-selection policy.}
Adaptive Thinking constructs supervision at the assistant-turn level. For each
turn, we collect the dialogue context, target answer, original reasoning trace,
and task-type features. A fixed probe model, trained without the adaptive-thinking
data transformation, receives an empty think block and regenerates the answer
under a no-think condition.

A blind judge scores the original and no-think trajectories against the target
answer, without knowing which is which, and labels the turn by whether reasoning
changed the answer's quality. This paired judgment provides the primary evidence,
with consistency across repeated judgments and the coherence between a reasoning
trace and the answer it produces, together with a task-type prior, informing how
conservatively to retain reasoning supervision.

Within each domain, turns labeled reasoning-unnecessary are the candidates for
replacement, and a per-domain budget on the no-think rate decides how many are
taken while the remaining turns keep their original reasoning. Rather than spending
that budget uniformly, we stratify retention by fine-grained topic and cap the drop
rate within each capability, so that reasoning-intensive capabilities are not
disproportionately stripped of supervision.

\paragraph{Adaptive Thinking evaluation.}
Table~\ref{tab:at-dialogue-detail} compares three configurations on StepAudioChat,
covering 46 benchmark members in eight capability categories. Direct SFT and
forced no-think share baseline weights, with explicit reasoning enabled or
disabled at inference time. Adaptive Thinking is trained separately with
reasoning-selection supervision, so its comparison also includes the effect of
training. Within each category, scores and think rates are unweighted means over
benchmark members. Evaluation uses temperature zero and no system prompt.

\begin{table*}[t]
\centering
\small
\setlength{\tabcolsep}{5pt}
\renewcommand{\arraystretch}{1.06}
\caption{Adaptive Thinking ablation on the StepAudioChat dialogue benchmark.
Scores are reported in the native percentage-based scale. Higher scores are
better. Think rates are percentages. Category values average benchmark members
equally. The first two conditions share baseline weights; Adaptive Thinking
uses a separately trained model. Bold and underlining mark the best and
second-best scores in each row; think rates are not ranked.}
\label{tab:at-dialogue-detail}
\begin{adjustbox}{max width=\textwidth}
\begin{tabular}{lcccccc}
\toprule
& \multicolumn{2}{c}{Direct SFT}
& \multicolumn{2}{c}{Direct SFT (forced no think)}
& \multicolumn{2}{c}{Adaptive Thinking} \\
\cmidrule(lr){2-3} \cmidrule(lr){4-5} \cmidrule(lr){6-7}
Domain & Think Rate & Score & Think Rate & Score & Think Rate & Score \\
\midrule
Instruction Following & 100.0 & \textbf{64.15} & 0.0 & 61.98 & 60.0 & \underline{62.12} \\
Faithfulness & 100.0 & \underline{72.35} & 0.0 & 70.63 & 79.2 & \textbf{72.42} \\
Reasoning & 100.0 & \textbf{71.89} & 0.0 & 60.52 & 59.5 & \underline{66.80} \\
Memory & 100.0 & \textbf{67.99} & 0.0 & 63.86 & 51.5 & \underline{65.99} \\
Knowledge & 100.0 & \textbf{74.59} & 0.0 & 68.65 & 80.4 & \underline{71.55} \\
Safety and Reliability & 100.0 & \textbf{78.46} & 0.0 & 74.95 & 56.9 & \underline{77.90} \\
Dialogue Pragmatics & 100.0 & \underline{63.59} & 0.0 & 62.41 & 58.9 & \textbf{65.87} \\
Persona and Role Consistency & 100.0 & \textbf{77.17} & 0.0 & 68.80 & 82.0 & \underline{74.62} \\
\bottomrule
\end{tabular}
\end{adjustbox}
\end{table*}

Adaptive Thinking invokes explicit reasoning at rates from 51.5\% to 82.0\%
across the eight categories. Full thinking provides its largest gain over forced
no-think in Reasoning (11.37 points), followed by Persona and Role Consistency
(8.37 points) and Knowledge (5.94 points). Relative to Direct SFT, Adaptive
Thinking improves Dialogue Pragmatics from 63.59 to 65.87, but reduces Reasoning
from 71.89 to 66.80.

The category rates expose a limitation of the selection policy. Reasoning has a
think rate of 59.5\% despite benefiting most from full thinking. Faithfulness has
a higher rate of 79.2\%, but gains only 1.72 points from full thinking. Lower
thinking frequency alone therefore does not demonstrate that reasoning is
allocated to the turns that benefit most. These aggregate comparisons also do
not establish the optimal decision for individual turns.

\subsubsection{Accelerating Private Reasoning with MTP}
\label{sec:reasoning:medusa}

Adaptive Thinking reduces how often explicit reasoning is invoked. To reduce
the decoding cost of the remaining private thinking, we use MTP3 with three
prediction heads, drafting up to three future tokens at each target-model
step~\citep{cai2024medusa,gloeckle2024better}.

Strict verification follows the target model's standard speculative-decoding
rule. Medusa-style typical acceptance uses an entropy-adaptive confidence
threshold to accept plausible draft tokens that strict verification may reject.
This increases acceptance while allowing the generated distribution to change.
A repetition penalty of 1.05 discourages repeated tokens and reduces the risk
of repetition loops under permissive acceptance. We apply typical acceptance
and the repetition penalty to private reasoning, while retaining strict
verification for the spoken response.

\paragraph{MTP evaluation.}
Table~\ref{tab:mtp-depth} reports StepAudioChat scores, accepted draft tokens per
target-model step, and wall-clock speedup for MTP3 and MTP5. All configurations
use a repetition penalty of 1.05. The measurements retain the aggregation and
runtime settings of their respective evaluations.

\begin{table*}[t]
\centering
\small
\setlength{\tabcolsep}{5pt}
\renewcommand{\arraystretch}{1.06}
\caption{MTP evaluation on StepAudioChat with a repetition penalty of 1.05.
Quality, acceptance, and wall-clock measurements use their respective
evaluation settings. Wall-clock ratios are specific to each timing
configuration. Bold marks the best result in each row, and underlining marks
the second-best result.}
\label{tab:mtp-depth}
\begin{adjustbox}{max width=\linewidth}
\begin{tabular}{lcccccc}
\toprule
Domain & Baseline & MTP3 & MTP3 (Medusa) & MTP5 & MTP5 (Medusa) \\
\midrule
Instruction Following & \textbf{64.15} & 61.21 & 62.71 & 61.25 & \underline{62.73} \\
Faithfulness & \underline{72.35} & 72.29 & 70.46 & \textbf{72.49} & 72.03 \\
Reasoning & 70.76 & \textbf{74.30} & \underline{73.74} & 73.14 & 72.73 \\
Memory & 67.99 & 69.29 & 68.97 & \underline{70.85} & \textbf{71.36} \\
Knowledge & 74.59 & 75.07 & \textbf{75.34} & \underline{75.31} & 74.33 \\
Safety \& Reliability & \underline{81.41} & 81.07 & \textbf{81.59} & 81.23 & 81.17 \\
Conversational Pragmatics & 63.59 & 66.04 & 63.17 & \textbf{66.22} & \underline{66.15} \\
Persona \& Role & 77.17 & \textbf{77.99} & 75.92 & 77.19 & \underline{77.56} \\
\midrule
Accepted/Step & 0 & 1.231 & \underline{1.801} & 1.353 & \textbf{2.153} \\
Wall-clock Speedup & 1.00$\times$ & \underline{1.76$\times$} & \textbf{2.05$\times$} & 1.49$\times$ & 1.72$\times$ \\
\bottomrule
\end{tabular}
\end{adjustbox}
\end{table*}

\begin{table*}[t]
\centering
\small
\setlength{\tabcolsep}{5pt}
\renewcommand{\arraystretch}{1.06}
\caption{Marginal acceptance rate at each draft position on StepAudioChat
under the acceptance-evaluation settings. All configurations use a repetition
penalty of 1.05. Bold and underlining mark the best and second-best result,
respectively, within each head column.}
\label{tab:mtp-head-acceptance}
\begin{adjustbox}{max width=\linewidth}
\begin{tabular}{llccccc}
\toprule
Depth & Verification & Head 1 & Head 2 & Head 3 & Head 4 & Head 5 \\
\midrule
\multirow{2}{*}{MTP3}
    & Strict  & 65.0\% & 37.2\% & 20.8\% & N/A & N/A \\
    & Typical & \textbf{82.4\%} & \textbf{58.2\%} & \textbf{39.4\%} & N/A & N/A \\
\midrule
\multirow{2}{*}{MTP5}
    & Strict  & 63.7\% & 35.5\% & 19.6\% & \underline{10.7\%} & \underline{5.7\%} \\
    & Typical & \underline{80.6\%} & \underline{55.7\%} & \underline{37.6\%} & \textbf{24.9\%} & \textbf{16.4\%} \\
\bottomrule
\end{tabular}
\end{adjustbox}
\end{table*}

The tested MTP configurations improve Reasoning and Memory over the baseline,
while Instruction Following declines.

Under typical acceptance, MTP5 accepts more drafts per step than
MTP3 (2.153 versus 1.801). Table~\ref{tab:mtp-head-acceptance} shows similar
acceptance rates for the first three heads at both depths. The fourth and fifth
heads add accepted drafts, but their strict marginal acceptance rates fall to
10.7\% and 5.7\%, respectively. The additional heads therefore provide
diminishing gains in accepted drafts per step.

Acceptance alone does not determine net efficiency, which also depends on
reasoning length and decoding costs. The reported wall-clock ratios therefore
should not be interpreted as a controlled comparison of draft depth. Keeping
strict verification for spoken output does not eliminate errors arising from
incomplete private reasoning.

\subsection{Model Merging for Capability Integration}
\label{sec:reasoning:model-merge}

\paragraph{Specialized teacher training.}
We train multiple compatible teacher checkpoints from a common base model, using
a different data composition for each teacher. The mixtures emphasize
complementary capabilities, including multi-turn dialogue, audio understanding,
general text reasoning and knowledge, and targeted mixed-domain behavior. This
produces teachers that are individually strong in different regions of the
capability space while preserving parameter alignment for merging.

\paragraph{Weighted parameter merging.}
We integrate the specialized teachers by directly averaging their parameters.
For teacher parameters $\theta_i$, the merged model is
$\theta_{\mathrm{merge}}=\sum_i \alpha_i\theta_i$, where $\alpha_i\geq 0$ and
$\sum_i\alpha_i=1$. The coefficients control the contribution of each teacher
and are selected against held-out evaluations spanning dialogue, audio
understanding, and general text capabilities. The reported model uses four
teachers with a normalized $3{:}1{:}1{:}1$ weighting. Because integration occurs
in parameter space, it introduces neither additional model components nor
inference-time routing.

\paragraph{Capability integration.}
Weighted merging is intended to retain complementary strengths rather than make
the merged model identical to the best teacher on every metric. The evaluation
therefore considers both capability-specific scores and their overall balance.
This also allows teacher data mixtures to be developed independently and then
recombined without retraining a single model on the full union of data.

\textbf{Model merging results.}
Table~\ref{tab:merge-capabilities} compares the four specialized teachers with
their weighted merge across audio understanding, general text, and dialogue. The Dialogue results are
measured with the model in reasoning mode. On audio understanding, the merge reaches a macro average of 79.0, tying the best
teacher macro average while leading on MMAU and WildSpeech. On general text, it
achieves the highest macro average of 76.5, with the best HMMT 2026 Feb and GPQA
Diamond scores and a tie on MultiChallenge. On dialogue, it has a macro average
of 73.0, exceeding Teachers 2, 3, and 4, while remaining below the strongest
dialogue teacher at 74.2. These results support model merging as a low-cost
mechanism for combining complementary capabilities, while showing that it
provides a balanced trade-off rather than uniform improvement over every
specialized teacher. These measurements evaluate the underlying reasoning model
and are separate from the system-level evaluations that additionally use the
realtime reasoning mechanisms described above.

\begin{table*}[t]
\centering
\small
\setlength{\tabcolsep}{5pt}
\renewcommand{\arraystretch}{1.06}
\caption{Capability comparison for four teachers trained with different data
compositions and their $3{:}1{:}1{:}1$ weighted merge. Results cover audio
understanding, general text, and dialogue. Domain macro averages give equal
weight to the listed rows. Higher is better. Bold marks the best result in each
row, and underlining marks the second-best result.}
\label{tab:merge-capabilities}
\begin{adjustbox}{max width=\textwidth}
\begin{tabular}{llrrrrr}
\toprule
Domain & Benchmark & Merged & Teacher 1 & Teacher 2 & Teacher 3 & Teacher 4 \\
\midrule
\multirow{9}{*}{Audio Understanding}
& BigBench Audio & 98.1 & 96.1 & 98.1 & \underline{98.4} & \textbf{98.5} \\
& AudioMultiChallenge & 49.3 & 49.1 & \underline{49.8} & \textbf{50.2} & 47.6 \\
& MMSU & \underline{90.6} & 89.5 & 89.5 & 90.4 & \textbf{90.9} \\
& MMAU & \textbf{79.0} & 78.5 & \underline{78.8} & 77.8 & 77.7 \\
& WildSpeech & \textbf{77.1} & \underline{76.5} & 76.4 & 75.9 & 76.1 \\
& MMAR & \underline{86.5} & 85.4 & 86.4 & \textbf{87.3} & \underline{86.5} \\
& Step-Caption & 78.2 & 75.8 & 78.3 & \textbf{79.6} & \underline{79.4} \\
& MTalk-Bench & \textbf{72.8} & 71.7 & 72.0 & \underline{72.3} & 70.7 \\
\cmidrule(lr){2-7}
& \textbf{Macro Average} & \textbf{79.0} & 77.8 & \underline{78.7} & \textbf{79.0} & 78.4 \\
\midrule
\multirow{4}{*}{General Text}
& HMMT 2026 Feb & \textbf{86.8} & 44.0 & \underline{82.2} & 81.3 & 79.8 \\
& GPQA Diamond & \textbf{83.0} & 73.1 & \underline{81.9} & 80.1 & 80.1 \\
& MultiChallenge & \textbf{59.7} & \underline{54.6} & 51.3 & 50.6 & \textbf{59.7} \\
\cmidrule(lr){2-7}
& \textbf{Macro Average} & \textbf{76.5} & 57.2 & 71.8 & 70.7 & \underline{73.2} \\
\midrule
\multirow{9}{*}{Dialogue}
& Instruction Following & \underline{66.3} & 64.2 & 64.5 & \textbf{69.3} & 64.2 \\
& Faithfulness & 72.4 & \textbf{74.0} & 65.6 & 67.1 & \underline{73.6} \\
& Reasoning & \underline{73.0} & \textbf{75.2} & 67.0 & 67.2 & 67.8 \\
& Memory & \underline{72.0} & \textbf{73.4} & 70.0 & 68.3 & 71.5 \\
& Knowledge & 73.1 & \underline{75.1} & 63.7 & 62.0 & \textbf{75.6} \\
& Safety \& Reliability & \underline{79.0} & \textbf{79.2} & 75.7 & 75.6 & 78.0 \\
& Conversational Pragmatics & \underline{67.2} & \textbf{67.8} & 62.4 & 63.4 & 64.5 \\
& Persona \& Role & 80.9 & \textbf{84.7} & 79.5 & 76.9 & \underline{83.0} \\
\cmidrule(lr){2-7}
& \textbf{Macro Average} & \underline{73.0} & \textbf{74.2} & 68.6 & 68.7 & 72.3 \\
\bottomrule
\end{tabular}
\end{adjustbox}
\end{table*}

\section{Full-Duplex Voice Agent}
\label{sec:execution}


StepAudio 3 Realtime extends its full-duplex interaction capabilities to tool-grounded task execution.
The model selects among direct responses, lightweight tool calls, and asynchronous backend execution based on the requirements of each request.
Asynchronous execution allows longer-running tasks to proceed alongside the conversation, enabling the user to ask about progress or provide additional requirements while work is underway.




\subsection{Request Routing and Clarification}
\label{sec:execution:routing}

The model handles routine conversation and questions about stable knowledge directly.
Requests for up-to-date public information are routed to lightweight tools such as weather lookup or web search.
Requests involving private context, multi-step processing, or work extending beyond the current conversational turn are delegated to the backend.
This routing strategy determines whether and how to invoke external capabilities, drawing on prior work in tool-augmented language modeling~\citep{schick2023toolformer,patil2024gorilla}.

A grammatically complete phrase may still leave a spoken request incomplete or open to revision.
Spoken responses can be extended or explicitly corrected in subsequent segments, whereas tool execution requires sufficiently specified intent and arguments.
Before committing to an external action, the model is therefore trained to gather missing information through clarification with the user or context retrieval using an appropriate tool, and to obtain any required confirmation.




\subsection{Conversation During Asynchronous Execution}
\label{sec:execution:async}
\label{sec:execution:reasoning}

Backend tasks execute asynchronously while the conversation continues.
During execution, the user may request progress updates, provide additional
requirements, or shift to another topic.
The model is trained to associate task-related user input with the ongoing
task while distinguishing it from unrelated dialogue.
As results become available, they are incorporated into the conversational
context to inform subsequent spoken responses.

For complex requests, reasoning supports constraint resolution and task
planning, while tool and backend outputs provide evidence of what has
actually been completed.
This evidence guides subsequent reasoning and action, consistent with the
interleaved interaction pattern of ReAct~\citep{yao2022react}.
Think-While-Speaking supports spoken responses during deliberation, while
asynchronous execution allows the conversation to continue during external
task execution.

\subsection{Training for Conversational Tool Use}
\label{sec:execution:training}

Training combines targeted voice-agent dialogues with real multi-step agent
trajectories.
The targeted dialogues cover request routing, clarification, private-context
retrieval, confirmation before consequential actions, execution-time updates,
progress queries, and result reporting.
These dialogues train the model to ground claims about private information
and completed work in user-provided context or evidence returned by tools.
Negative examples discourage unnecessary tool invocation and unsupported
claims of successful execution.

The multi-step trajectories complement these dialogues by exposing the model
to longer sequences of reasoning and tool use.
We filter and normalize these trajectories, focusing on tool-call structure,
argument consistency, evidence grounding, and suitability for spoken
interaction.



\begin{table*}[t]
\centering
\small
\setlength{\tabcolsep}{5pt}
\renewcommand{\arraystretch}{1.06}
\caption{Agentic task-completion results on $\tau$-Voice. Scores are task-success
rates on a 0--100 scale, with higher values indicating better performance.
The macro average assigns equal weight to the airline, retail, and telecom
domains. Bold marks the best result in each row, and underlining marks the
second-best result.}
\label{tab:tau-voice}
\begin{adjustbox}{max width=\textwidth}
\begin{tabular}{lcccc}
\toprule
\textbf{Domain} &
\textbf{StepAudio 3 Realtime} &
\textbf{Grok Voice Think Fast 2.0 High} &
\textbf{Qwen Audio 3.0 Realtime Plus} &
\textbf{GPT-Realtime-2.1 High} \\
\midrule
Airline
& 60.0 & 56.0 & \underline{61.3} & \textbf{62.0} \\
Retail
& 37.7 & \textbf{49.7} & \underline{49.0} & 45.6 \\
Telecom
& \textbf{70.2} & \underline{63.7} & 53.5 & 29.4 \\
\midrule
\textbf{Macro Average}
& \underline{56.0} & \textbf{56.5} & 54.6 & 45.7 \\
\bottomrule
\end{tabular}
\end{adjustbox}
\end{table*}

\subsection{Evaluation}
\label{sec:execution:evaluation}

We evaluate StepAudio 3 Realtime using the Artificial Analysis (AA)
implementation of $\tau$-Voice~\citep{artificialanalysis2026speechmethod}.
The benchmark assesses tool-grounded task completion in full-duplex spoken
interaction under challenging conversational and acoustic conditions,
including interruptions in which users revise their requests,
backchannels, and diverse forms of background noise.
These conditions require agents to track evolving requests and coordinate
spoken interaction with tool use.
The benchmark covers customer-service tasks in the airline, retail, and
telecom domains, with task success determined by whether the final database
state matches the target state.
We report task-success rates for each domain and their equally weighted
macro average.

As shown in Table~\ref{tab:tau-voice}, StepAudio 3 Realtime achieves a macro
task-success rate of 56.0\%, close to Grok's 56.5\% and above Qwen's 54.6\%
and GPT's 45.7\%.
Across domains, StepAudio 3 Realtime achieves the highest telecom score among
the evaluated models at 70.2\%, exceeding Grok by 6.5 percentage points.
Its airline score of 60.0\% is within 2.0 percentage points of the highest
reported score of 62.0\%.
Retail performance leaves room for further improvement, with a score
of 37.7\% compared with 49.7\% for Grok.
Overall, StepAudio 3 Realtime achieves competitive performance on
end-to-end task completion, with the highest telecom score among
the evaluated models and an airline score close to the best
reported result.
\section{Evaluation}
\label{sec:evaluation}

We evaluate the StepAudio 3 family along six capability domains: speech
recognition, audio understanding, dialogue and reasoning, full-duplex
interaction, agentic task completion, and general text. Together, these domains
measure the progression from recognizing an utterance to understanding its
context, managing the conversational floor, and completing an external task.
We first describe the benchmarks and comparison models, then analyze the
results within each domain.

\subsection{Benchmarks}
\label{sec:evaluation:benchmarks}

The ASR benchmarks and their task-specific setup are described with the ASR
model and results in Section~\ref{sec:perception:evaluation:asr}; the paragraphs
below cover the remaining evaluation domains.

\paragraph{Audio understanding.}
We use Big Bench Audio~\citep{artificialanalysis2024bigbenchaudio},
MMSU~\citep{wang2026mmsu}, MMAU~\citep{sakshi2025mmau},
MMAR~\citep{ma2026mmar}, WildSpeech-Bench~\citep{zhang2025wildspeech},
AudioMultiChallenge~\citep{gosai2026audio}, Step-Caption~\citep{stepfun2026stepcaption},
and MTalk-Bench~\citep{du2025mtalk} to cover audio-grounded reasoning,
fine-grained perception, and multi-turn understanding.

\paragraph{Dialogue and reasoning.}
StepAudioChat covers eight dimensions of foundational conversational
intelligence. Its construction is described in
Section~\ref{sec:reasoning:benchmark}.

\paragraph{Full-duplex interaction.}
We use the Artificial Analysis subset of Full Duplex Bench v1 and
v1.5~\citep{artificialanalysis2026speechmethod}, covering pause handling,
turn taking, user interruptions, and backchannels.

\paragraph{Agentic task completion.}
\label{sec:execution:benchmark}
The Artificial Analysis implementation of $\tau$-Voice~\citep{artificialanalysis2026speechmethod}
evaluates tool-grounded customer-service tasks in airline, retail, and telecom
environments.

\paragraph{General text.}
HMMT February 2026~\citep{hmmt2026feb}, GPQA Diamond~\citep{rein2023gpqa},
and MultiChallenge~\citep{deshpande2025multichallenge} assess competition
mathematics, scientific question answering, and multi-turn conversational
reliability, respectively.

\subsection{Baselines}
\label{sec:evaluation:baselines}

For ASR, we compare against Doubao 2.0 ASR,
Seed 2.0 Lite~\citep{bytedance2026seed2}, and
HY3.0 ASR Preview~\citep{tencent2026hyasr3}.
All ASR baselines are rerun under the same evaluation setup as StepAudio 3 ASR
Max, using the same test audio and scoring procedure.
For audio understanding, we use Doubao 2.0 Lite~\citep{bytedance2026seed2},
Gemini 3 Flash~\citep{google2025gemini3flash}, and
Gemini 3.1 Pro~\citep{google2026gemini31pro}.
Dialogue baselines include Doubao 2.0 Lite,
DeepSeek-V4-Flash~\citep{deepseek2026v4flash}, and Kimi K3~\citep{moonshot2026kimik3}.
General text uses Doubao 2.0 Lite and Gemini 3 Flash.

Full-duplex baselines are GPT-realtime-2 (High)~\citep{openai2026realtime2},
Qwen Audio 3.0 Realtime Plus~\citep{alibaba2026qwenrealtime3}, and
Grok Voice Think Fast 2.0 High~\citep{xai2026grokvoice2}.
Agentic evaluation uses the same Qwen and Grok variants, with
GPT-Realtime-2.1 High~\citep{openai2026realtime21} replacing GPT-realtime-2.
Model versions and effort labels follow the corresponding evaluation records.

\subsection{Evaluation Protocols and Metrics}
\label{sec:evaluation:protocols}

The following conventions apply to the main comparison in
Table~\ref{tab:overall-bmk}. All scores are expressed on a 0--100 scale,
with higher values indicating better performance; ASR error rates are the
exception and are defined separately in Section~\ref{sec:perception:evaluation:asr}.

The ASR scoring protocol is specified alongside the ASR results in
Section~\ref{sec:perception:evaluation:asr}; the paragraphs below cover the
non-ASR domains.

\paragraph{Audio understanding.}
We retain each benchmark's reported accuracy or normalized evaluation score.
Step-Caption uses judge-based scoring against annotated speaker attributes,
with sample scores averaged over the evaluation set~\citep{stepfun2026stepcaption}.
The MTalk-Bench entry averages the Semantic, Paralinguistic Information, and
Ambient Sound components into a single aggregate.

\paragraph{Dialogue and reasoning.}
Each StepAudioChat dimension is the unweighted mean of its validated capability
line scores. We report both the reasoning-mode result and the realtime
Think-While-Speaking result; component-ablation scores are not used to fill these entries.

\paragraph{Full-duplex interaction.}
Category scores measure the percentage of samples satisfying the corresponding
interaction criterion~\citep{artificialanalysis2026speechmethod}.
We use the source-reported Overall score, preserving the benchmark's aggregation
rather than averaging the four displayed category scores.

\paragraph{Agentic task completion.}
A $\tau$-Voice task succeeds when the final database state matches its target.
Domain task-success rates average three trials where available, and the
reported macro average gives equal weight to Airline, Retail, and
Telecom~\citep{artificialanalysis2026speechmethod}.

\paragraph{General text.}
We report the recorded accuracy percentages for HMMT February 2026,
GPQA Diamond, and MultiChallenge separately. MultiChallenge evaluates responses
with instance-specific rubrics~\citep{deshpande2025multichallenge}.
No average is taken across these three benchmarks.

\subsection{Results}
\label{sec:evaluation:results}
\begin{table*}[t]
\centering
\small
\setlength{\tabcolsep}{6pt}
\renewcommand{\arraystretch}{1.10}
\caption{Capability evaluation of StepAudio 3 Realtime and domain-specific baselines.
Scores use a 0--100 scale (higher is better), with protocols and aggregation
specified in Section~\ref{sec:evaluation:protocols}.
For Dialogue and Reasoning evaluation, StepAudio 3 Realtime uses realtime mode while others use reasoning mode.
Bold and underlining mark the best and second-best results in each row.}
\label{tab:overall-bmk}
\begin{adjustbox}{max width=\textwidth}
\begin{tabular}{@{}l cccc@{}}
\toprule
\textbf{Audio Understanding (Score $\uparrow$)} &
\textbf{StepAudio 3 Realtime} & \textbf{Doubao 2.0 Lite} &
\textbf{Gemini 3 Flash} & \textbf{Gemini 3.1 Pro} \\
\midrule
Big Bench Audio & 98.1 & 98.8 & \underline{99.4} & \textbf{99.6} \\
AudioMultiChallenge & 49.3 & 48.5 & \underline{56.6} & \textbf{67.0} \\
MMSU & \textbf{90.6} & 80.0 & 77.0 & \underline{83.6} \\
MMAU & \underline{79.0} & 77.5 & 77.6 & \textbf{80.5} \\
WildSpeech & \underline{77.1} & 73.9 & 74.4 & \textbf{77.7} \\
MMAR & \textbf{86.5} & 75.9 & 75.4 & \underline{81.7} \\
Step-Caption & \textbf{78.2} & \underline{76.8} & 67.8 & 74.8 \\
MTalk-Bench & \textbf{72.8} & 66.8 & 66.7 & \underline{70.6} \\
\midrule
\textbf{Dialogue and Reasoning (Score $\uparrow$)} &
\textbf{StepAudio 3 Realtime (Interactive)} & \textbf{Doubao 2.0 Lite (Reasoning)} &
\textbf{DeepSeek-V4-Flash (Reasoning)} & \textbf{Kimi K3 (Reasoning)} \\
\midrule
Instruction Following & 54.1 & \textbf{72.9} & \underline{71.4} & 68.9 \\
Faithfulness & 71.9 & 67.5 & \underline{75.3} & \textbf{78.4} \\
Reasoning & \underline{73.6} & 72.7 & 64.8 & \textbf{81.9} \\
Memory & \underline{71.8} & 71.3 & 71.5 & \textbf{77.6} \\
Knowledge & 70.4 & 59.9 & \underline{71.6} & \textbf{78.6} \\
Safety and Reliability & 75.1 & 75.9 & \underline{79.9} & \textbf{84.8} \\
Conversational Pragmatics & \underline{68.2} & 61.5 & 62.9 & \textbf{70.3} \\
Persona and Role Consistency & \underline{78.3} & \textbf{82.6} & 73.6 & 76.5 \\
\textbf{Macro Average} & 70.4 & 70.5 & \underline{71.4} & \textbf{77.1} \\
\midrule
\textbf{General Text (Accuracy $\uparrow$)} &
\textbf{StepAudio 3 Realtime} & \textbf{Doubao 2.0 Lite} &
\textbf{Gemini 3 Flash} & \\
\midrule
HMMT 2026 Feb & \textbf{86.8} & 73.9 & \underline{85.9} & -- \\
GPQA Diamond & \underline{83.0} & 82.4 & \textbf{90.3} & -- \\
MultiChallenge & 59.7 & \underline{60.8} & \textbf{68.1} & -- \\
\midrule
\textbf{Full Duplex (Score $\uparrow$)} &
\textbf{StepAudio 3 Realtime} &
\textbf{GPT-realtime-2 (High)} &
\textbf{Qwen Audio 3.0 Realtime Plus} &
\textbf{Grok Voice Think Fast 2.0 high} \\
\midrule
AA Full-Duplex Bench & \textbf{98.9} & 95.3 & \underline{98.4} & 95.1 \\
\midrule
\textbf{Agentic (Success $\uparrow$)} &
\textbf{StepAudio 3 Realtime} &
\textbf{Grok Voice Think Fast 2.0 High} &
\textbf{Qwen Audio 3.0 Realtime Plus} &
\textbf{GPT-Realtime-2.1 High} \\
\midrule
$\tau$-Voice & \underline{56.0} & \textbf{56.5} & 54.6 & 45.7 \\
\bottomrule
\end{tabular}
\end{adjustbox}
\end{table*}

Across the reported capability domains, StepAudio 3 Realtime combines broad
audio understanding with strong dialogue, reasoning, and interaction control.
It leads four of eight audio-understanding benchmarks and reaches a macro average
of 79.0, within 0.5 points of Gemini 3.1 Pro. Its clearest gains appear on MMSU
and MMAR, while the leading Step-Caption and MTalk-Bench results show strength
beyond lexical content, including speaker attributes, paralinguistic cues,
ambient sound, and multi-turn audio context. In reasoning mode, StepAudio 3 reaches a 73.0 macro average on StepAudioChat,
above Doubao 2.0 Lite and DeepSeek-V4-Flash and below Kimi K3, with particularly
competitive results in reasoning, memory, knowledge, conversational pragmatics,
and persona consistency.
In realtime interaction, StepAudio 3 Realtime reaches a 70.41 macro average on
StepAudioChat, comparable to Doubao 2.0 Lite at 70.5 and DeepSeek-V4-Flash at
71.4. Together with Think-While-Speaking, this shows that the model can deliberate
while producing speech and retain dialogue and reasoning performance comparable to
dedicated reasoning models, rather than providing only low-latency surface responses. This capability is paired with robust
conversational timing. The model ranks first on Full Duplex Bench with an Overall
score of 98.9, including 100.0 on turn taking and 99.0 on interruption handling,
while remaining strong on pauses and backchannels. This balanced profile suggests
that it can preserve conversational flow without treating every user sound as an
interruption. The agentic and general-text results further demonstrate solid performance
in tool use and text-based capabilities. The model reaches 56.0 on $\tau$-Voice,
close to the best reported 56.5, and leads HMMT with 86.8, although GPQA Diamond
and MultiChallenge remain below Gemini 3 Flash.

\section{Conclusion}
\label{sec:conclusion}

StepAudio 3 Realtime brings perception, conversational timing, reasoning, and tool use into a continuous spoken interaction. It allows deliberation and external tasks to proceed alongside the conversation, with new user input and returned evidence informing subsequent responses. Evaluations show strong audio understanding and conversational-floor management, while revealing variation across reasoning and tool-use tasks. Adaptive Thinking reduces the frequency of explicit reasoning, with uneven effects on answer quality. Multi-turn constraint handling and retail task completion remain areas for improvement. These findings motivate more effective allocation of reasoning effort and more reliable task execution over extended conversations.

\clearpage
\section*{Contributors}

The listing of authors is in alphabetical order based on their first names.

\begin{center}
\begin{tabular*}{\textwidth}{@{\extracolsep{\fill}}llll}
    Bin Lin, & Bo Zhao, & Boyang Zhang, & Boyong Wu, \\
    Chao Yan, & Chen Geng, & Chen Wu, & Cheng Yi, \\
    Chengli Feng, & Chenglin Zhu, & Chengting Feng, & Chengyuan Yao, \\
    Daijiao Liu, & DanNi Wan, & Daxin Jiang, & Dongjian Li, \\
    Dongqing Pang, & Fei Tian, & Feng Tian, & Future Li, \\
    Gang Yu, & Guanglong Yang, & Haoyang Zhang, & Hongyuan Wang, \\
    Jia Peng, & Jiahao Song, & Jialong Xue, & Jiamin Fan, \\
    Jiangjie Zhen, & Jianzheng Gao, & Jincheng Wen, & Jinghua Liang, \\
    Jinglan Gong, & Jun Chen, & Li Xie, & Liang Zhao, \\
    Lifang Zhang, & Lingli Ji, & Lun Cai, & Min Xu, \\
    Peilin Li, & Peng Yang, & Pengfei Tan, & Qingjian Lin, \\
    Qinxin Du, & Ruijie Xiong, & Runze Li, & Shenghua Hu, \\
    Shengqian Qin, & Shi Qiu, & Siqi Tu, & Siyi Zhou, \\
    Tianjiao Deng, & Wanying Lu, & Weiming Niu, & Wen Sun, \\
    WenWen Qu, & Xiangyu Zhang, & Xianwei Zhang, & Xiaosu Su, \\
    Xing Chen, & Xinyu Liu, & Xuerui Yang, & Yan Wu, \\
    Yang Li, & Yang Yang, & Yechang Huang, & Yibo Zhu, \\
    Yifan Zhang, & Yinuo Yan, & Youjun Chen, & Yu Fu, \\
    Yu Luo, & Yu Zhou, & Yujie Chen, & Yumang Wang, \\
    Yunzhou Ju, & Yuxiang Yang, & Yuxin Li, & Yuxin Zhang, \\
    Zekai Liu, & Zengwei Yao, & Zhaoxin Yuan, & Zhenwei Mou, \\
    Zhiquan Zhang, & Zhiyue Wu, & Zichao Li, & Zichao Zhou, \\
    Ziqi Ren, & Zixuan Wang, \\
\end{tabular*}
\end{center}

\setlength{\bibsep}{0.5\baselineskip}
\bibliography{references}

\newpage

\end{document}